\documentclass[prl,twocolumn,showpacs,superscriptaddress]{revtex4}

\usepackage{graphicx}
\usepackage{color}

\begin{document}

\title{Observation of correlations up to the micrometer scale in sliding charge-density waves}
             
\author{D. Le Bolloc'h}
\affiliation{Laboratoire de Physique des Solides (CNRS-UMR 8502), B{\^a}t. 510, Universit{\'e} Paris-sud, 91405 Orsay cedex, France}
\author{V.L.R. Jacques}
\affiliation{Laboratoire de Physique des Solides (CNRS-UMR 8502), B{\^a}t. 510, Universit{\'e} Paris-sud, 91405 Orsay cedex, France}
\author{N. Kirova}
\affiliation{Laboratoire de Physique des Solides (CNRS-UMR 8502), B{\^a}t. 510, Universit{\'e} Paris-sud, 91405 Orsay cedex, France}
\author{J. Dumas}
\affiliation{Institut N\'eel, CNRS/UJF, BP166 38042 Grenoble cedex 9, France}
\author{S. Ravy}
\affiliation{Synchrotron SOLEIL, L'Orme des merisiers, Saint-Aubin BP 48, 91192 Gif-sur-Yvette cedex, France}
\author{J. Marcus}
\affiliation{Institut N\'eel, CNRS/UJF, BP166 38042 Grenoble cedex 9, France}
\author{F. Livet}
\affiliation{LTPCM (CNRS-UMR 5614), ENSEEG-Domaine Universitaire, BP 75, 38402 Saint Martin d'H\`eres cedex, France}

\begin{abstract}
High-resolution coherent x-ray diffraction experiment has been performed on 
 the charge density wave (CDW) system 
K$_{0.3}$MoO$_3$. The $2k_F$ satellite reflection associated with
the CDW has been measured with respect to external dc currents. In the sliding regime,
the $2k_F$ satellite reflection displays
 secondary satellites along the chain axis which
corresponds to correlations up to the micrometer scale. This super long range order is
1500 times larger than the CDW period itself. This new type of electronic correlation
seems inherent to the collective dynamics of electrons in charge density wave systems.
Several scenarios are discussed.
\end{abstract}

\maketitle
 
Ferroelectrics, magnetic systems, liquid crystals,
spin and charge density waves may stabilize incommensurate
 structures with the underlying lattice\cite{cummins}. In most cases,
the stability of incommensurate modulations
 relies upon a coupling with the underlying lattice
or intrinsic defects.
The case of charge density waves is particular because
the electronic modulation is mainly driven by the electronic structure. 
The nesting of the Fermi surface 
stabilizes the modulation and the band filling fixes the wave vector of the modulation
at twice the Fermi wave vector $2k_F$, which may be incommensurate. 
The K$_{0.3}$MoO$_3$ blue bronze system, for example, stabilizes an incommensurate
 Charge Density Wave (CDW) modulation at 2$k_F$ =0.748$\pm$0.001\cite{JPP} along the chain axis.

This incommensurate 2$k_F$ wave vector has a well-known consequence: 
the invariance by translation
allows the electronic crystal to slide over the underlying lattice for
currents greater than a threshold value. 
The signature of this sliding motion has been
 mainly observed by transport measurements\cite{ReviewBB,grunermonceau}:
a large broad band noise is observed,
as well as periodic voltage oscillations\cite{fleming,hundlay}. 

Observing the collective dynamics of electrons by classical diffraction has
 always been a significant challenge.
The task is difficult in CDW systems 
because the CDWs domains are usually larger than the micrometer scale 
and any translational motion does not change the diffraction pattern.
Only few consequences of the sliding have been observed
 by using high-resolution x-ray diffraction, as for example, 
the loss of transverse coherence
of CDWs domains\cite{tamegai}, the contraction of the CDW close to electrical contacts\cite{brazo} 
or its rotation around a step\cite{thorne}.
By using high-resolution ${\it and}$ coherent x-ray diffraction, 
we show in this Letter a novel consequence of the sliding CDW and the incommensurate 2$k_F$ wave vector: 
 a super long range order, up to the micrometer scale, appears in the sliding regime of the CDW.
 This observation, as well as an earlier one reported in \cite{lebolloch1}, became possible only
 by using coherent micro diffraction.

\begin{figure}
\centering
\includegraphics[width=0.5\textwidth]{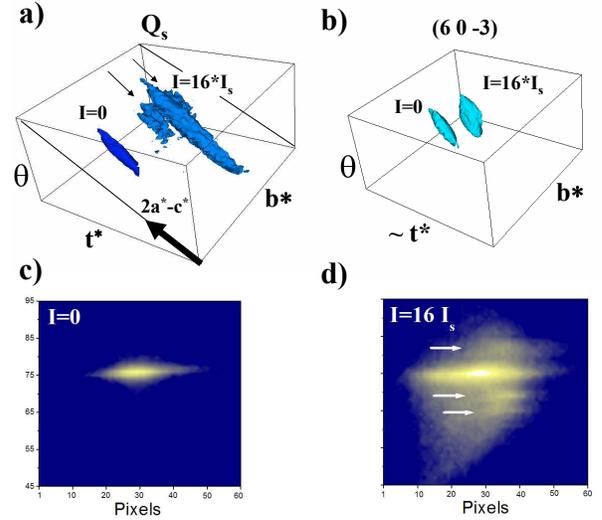}
\hspace{-4cm}\caption{\label{figure1} 3D diffraction patterns of a) the $Q_s=(6,\overline{0.252},\overline{3.5})$ satellite and b) the $(6,0,\overline{3})$ fundamental
 Bragg reflection for two current values at T=75K. Each isosurface has been fixed at $I_{max}$/18
 for the $(6,0,\overline{3})$ and $I_{max}$/7 for $Q_s$. For clarity, 
the reflections with and without current have been shifted along ${\bf b^*}$. 
The total size of the
boxes is 40*($\delta\theta=10^{-3}$ degree) (vertical axis) by 50*60 pixels (horizontal plane).
Each 3D acquisition lasted less than 15 mn.
2D patterns corresponding to integration of the fig 1a over the $\theta$ axis, without c) and with current d).
The secondary satellites are indicated by arrows. 
}
\end{figure}

We have studied the blue bronze which is a classic CDW system. 
This is a quasi-1D structure, made of clusters 
of 10 MoO$_{6}$ octahedra, forming chains along
the [010] direction (b=7.56 $\AA$), 
and layers along the [102] direction. 
The reciprocal vector ${\bf b^*}$  
 runs along the chains and ${\bf 2a^*-c^*}$ is perpendicular to the layers.
Let ${\bf u}({\bf r})={\bf u}_{0}\cos ({\bf q}_{c}.{\bf r}+\Phi({\bf r}))$ be 
 the periodic lattice distortion in
quadrature with the CDW, where
 ${\bf q}_c=(1, 2k_{F}, \overline{0.5})$ is defined as the wave vector
normal to the CDW wave fronts. At equilibrium, the CDW exhibits intrinsic defects like dislocations
\cite{lee3,lebolloch1} and displays its own vibration modes\cite{hennion,ravy1}. 
The dispersion curve of the CDW acoustic mode, the so-called phason mode, which is responsible for 
the sliding motion of the CDW as a whole, has been measured by inelastic 
neutron scattering\cite{hennion}.

We used the same high quality single crystal as in our previous experiment\cite{lebolloch1}. 
Its electrical resistance has been carefully measured before and after the experiment.
In both cases, a pronounced decrease of the differential resistance was observed at the threshold current,
accompanied by a rapid increase of the broad band noise, which is a signature 
of the sliding state. 
The transition temperature ($T_c$=180K) and the threshold current ($I_s$=1.2mA at 75K) 
remained unchanged after the experiment: x-rays did not alter the macroscopic CDW's properties.
No heating effect under the x-ray beam has been observed.

The coherent x-ray diffraction experiment has been performed at the 
ID01 beamline at the ESRF.
The 0.5$\times$2$\times$0.2mm$^3$ sample was mounted in a top-loading Cryostat cooled down to 75K with He exchange gas.
The sample was initially aligned with the {\bf b$^*$} axis vertical, and the 
$2{\bf a}^{*}-{\bf c}^{*}$ axis in the horizontal scattering plane. 
The patterns were recorded on a direct illuminated CCD
camera (22$\mu$m$\times$22$\mu$m pixel size) located 1.20m from the sample position.
 The beam quality and its transverse coherence length were tested
 by closing the entrance slit at $2\mu m \times 2\mu m$: the expected cross-like diffraction 
 pattern was observed with strong contrast of fringes\cite{lebolloch2}. 
 
 The degree of coherence can be estimated from the experimental setup around $10\%$\cite{livet2}.
 The penetration depth was $18\mu m^{-1}$ and the footprint of the xray beam on the sample's surface was $45 \mu m$
 (the incident angle equals $12.6^{o}$).
 In this study, the main interest of using a coherent beam is the small beam size
 and the excellent Q-resolution. 
A resolution of $\delta q=0.7\ 10^{-4} \AA^{-1}$ along ${\bf b^*}$ 
(i.e. $\delta q =0.8\ 10^{-4}$ in ${\bf b^*}$ units.)
 was achieved at 7.5 keV ($\lambda=1.65 \AA$) by using $10\mu m \times 10\mu m$
entrance slits.

\begin{figure}
\centering
\includegraphics[width=0.46\textwidth]{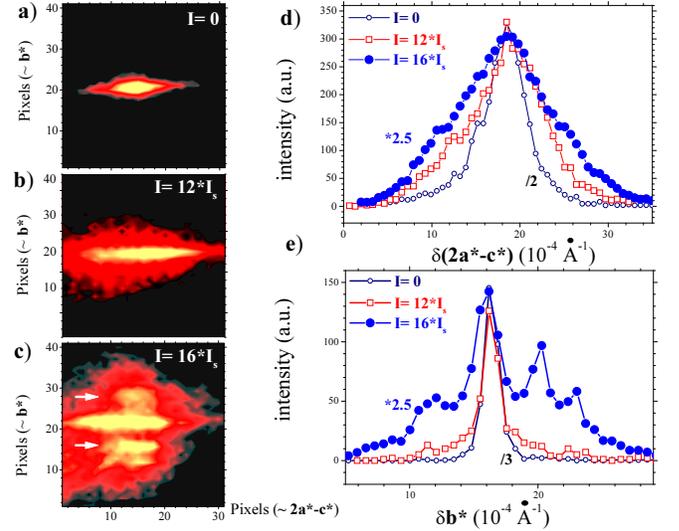}
\caption{\label{figure2} 
2D diffraction patterns (30*40 pixels) of the $2k_F$ reflection in the ($2{\bf a^*}-{\bf c^*}$;${\bf b^*}$) plane 
for I=0 mA (a), I=12 $I_s$ (b) and I=16 $I_s$ (c). 
The secondary satellites are indicated by arrows.
Corresponding profiles along ${\bf 2a^*}$-${\bf c^*}$ in d) and ${\bf b^*}$ in e).
The profiles have been rescaled.}
\end{figure}

The experiment consisted in recording the 2D diffraction patterns 
of the ${\bf Q_s}$=$(5,\overline{1},\overline{3})$+${\bf q}_c$ satellite reflection,
 and the $(6,0,\overline{3})$
fundamental Bragg peak, far from any electric contact. Both reflections have been measured 
successively after each current variation.
Due to the experimental geometry, the 2D reciprocal 
plane probed by the CCD at the satellite angle corresponds to the (${\bf b^*}$,${\bf t^*}$) plane, 
where ${\bf t^*}$ is the direction tilted by
$19.5^{o}$ from the ${\bf 2a^*}$-${\bf c^*}$ direction\cite{note7}.
Several CCD acquisitions have been recorded for different incident $\theta$ angles, 
with $\delta \theta=10^{-3}$ degree steps.
 This allows us to get the three dimensional intensity distribution of the reflections (see Fig.\ref{figure1}). 
 In this way, the behavior of the $2k_F$ reflection and the main Bragg reflection was observed 
 with respect to external current, 
especially along the chain direction ${\bf b^*}$ and the transverse direction
${\bf 2a^*}$-${\bf c^*}$. In order to get more intensity, 
we have integrated the volume of the Fig.\ref{figure1} over the $\theta$ axis. The corresponding 2D pattern
is displayed without current in Fig.\ref{figure1}c and with current in Fig.\ref{figure1}d.
The 2D patterns shown in Fig.\ref{figure2} correspond to a section of Fig.\ref{figure1} along the $2{\bf a^*}-{\bf c^*}$
direction through the $2k_F$ reflection.

First, the host lattice seems to be 
insensitive to the applied current: the $(6,0,\overline{3})$ fundamental Bragg peak remains unchanged under the applied current
within the experimental resolution (Fig.\ref{figure1}b). 
On the other hand, the CDW modulation displays original features for large enough currents.
Under current,
a broadening of the $2k_F$ reflection is observed along the ${\bf 2a^*}$-${\bf c^*}$ transverse direction  
which corresponds to decreasing CDW correlation lengths from $\xi_t=1.2\mu m$ at I=0mA,
 to $\xi_t=0.6\mu m$ at I=12 $I_s$ and to $\xi_t=0.4\mu m$ at I=16 $I_s$ (Fig.\ref{figure2}d).
The loss of the transverse order under external currents, along the softest direction of blue bronze\cite{note5},
has already been observed in several studies\cite{tamegai}.

The most striking feature appears along the ${\bf b^*}$ direction.
Without current, the $2k_F$  reflection displays a single peak, the width of which
corresponds to the entrance slit aperture: the CDWs domain is larger than $10\mu m$ along ${\bf b^*}$.
Above approximately 12 times the threshold current,  the 
$Q_s$ satellite starts to broaden along ${\bf b^*}$ (Fig.\ref{figure2}b).
This broadening increases continuously for larger currents. At
 a current 16 times larger than the threshold, several maxima located
  at regular positions along ${\bf b^*}$   are 
clearly distinguished from the 2$k_F$ reflection (see Fig.\ref{figure1}d).
All of them display approximately the same width, slightly larger than the $2k_F$ width without current. 
As shown in Fig.\ref{figure1}, these new intensity maxima correspond to well defined 3D peaks and will be called ${\it secondary\ satellite"}$ reflection in the following. 
These secondary satellite reflections are located at $(6,\overline{0.252}\pm n\delta q_s,\overline{3.5})$ with $\delta q_s=4.9\ 10^{-4}$ ${\bf b^*}$ units (see Fig.\ref{figure1}a,\ref{figure1}d, \ref{figure2}c and \ref{figure2}e). 
The reduced wave vector $\delta q_s$ leads to 
a period of L=${2\pi}/{\delta q_s}$=1.5$\mu$m. 

We did not increase the current further for fear of destroying the electrical contacts.
The appearance of secondary satellite reflections is reversible: 
after heating up the sample above $T_c$ under zero field and cooling down to 75K again, 
the same diffration pattern has been observed\cite{note6}. 
These secondary satellite reflections were also observed in a second high quality crystal.
Note the  strong intensity of higher-order satellites and
 the asymmetry of the profile in Fig.\ref{figure2}e\cite{note4}. 
 Fig.\ref{figure1}d also displays an asymmetric diffuse intensity along the transverse direction.
 This is surely due to inhomogeneities in the CDW current density within the probed volume.
 Those inhomogeneities have been observed by transport measurements in the blue bronze\cite{hundlay}. 
 We also noticed 
 the absence of any speckle neither on the 2$k_F$ reflection nor on secondary satellites, 
 despite the coherence properties of our x-ray beam. 
This last point will be brought up later.

Let us discuss the interpretation. First, the secondary satellites cannot be due to 
the fragmentation of CDW in isolated CDWs domains,
1.5 $\mu$m wide along ${\bf b^*}$. The truncation effect due to finite CDWs domain
would lead to very weak satellites with respect to the main $2k_F$ reflection. 
It can not be explained either by the presence
of a single CDW dislocation as it has been observed in blue bronze without current in \cite{lebolloch1}.
Indeed,  the distance between each fringe would be constant whatever the current and equal 
to the beam size, i.e. $\delta q={2\pi}/{10\mu m}=0.6 \ 10^{-4} \AA^{-1}$. In addition, the main $2k_F$ reflection
should disappear because of the presence of the dislocation. 
The last two points are in contradiction with our measurement.
Finally, phase shifts randomly distributed in the volume would lead to typical speckle patterns\cite{ravy2}
and not to regular satellite reflections as observed in Fig.\ref{figure1}d. 
 Note also that purely dynamical phenomena, like phase slips,
  would only have a negligible effect on the 2$k_F$ reflection.

We interpret the  experimental data as due to
a long range periodic structure which modulates the CDW state along ${\bf b^*}$.
This super long range order is surprising since
the period L is 1500 times larger than the CDW period along the chain 
direction ($\lambda_{CDW}\approx 4/3$ b=10.08 $\AA$). 
To our knowledge, long range electronic correlations up to the micrometer scale
 have never been observed in electronic systems.

The existence of this super long range order
 is a direct consequence of the incommensurability of the $2k_F$ wave vector 
and of the sliding motion of the electronic crystal.  
Several scenarios are briefly discussed here, based on static or dynamic phenomena:
a soliton lattice, an ordering of CDW dislocations or 
a new vibration mode induced by sliding.

 The observed periodic sequence of
satellites flanking the $2k_F$ reflection 
displays a remarkably high and slowly decaying intensity.
 It signifies the appearance under current of
non-harmonic periodic structure, modulating the CDW state
along ${\bf b^*}$. An amplitude modulation of the CDW would yield symmetrical secondary satellite reflections, 
that only structure factor or Debye-Waller effects could make asymmetric in intensity. On the other hand, a phase modulation would directly result in asymmetric secondary satellite reflections \cite{cummins,cowley}. This remark, along with the well-known fact that phase excitations cost much less energy that amplitude ones, lead us to consider this latter case in the following. 

Consider first a long range modulation provided by a static soliton
lattice. In blue bronzes, the CDW wave vector ($Q=0.748{\bf b^*}$) is
close to $0.750{\bf b^*}$ corresponding to the
quarterization along the chain. The actual
crystallographic space group requires an 8-order commensurability ($8{\bf q_c}$ is a
fundamental Bragg reflection) which locks the CDW phase
$\varphi$ at multiples of $\pi /4$ via the energy
$\sim-\alpha \cos(8\varphi)$\cite{anderson}.  Thus, the
 proximity to the
commensurability point results in a lattice of discommensurations,
 or  solitons with the phase increment $\Delta \varphi={\pi}/{4}$.
The soliton lattice is characterized by the soliton size $l$ and
the distance between the solitons $L$. The single soliton size l$\sim
(C/\alpha)^{1/2}$ is then determined jointly by the energy $\alpha$ and
the CDW longitudinal elastic modulus $C$. 
At low applied current
the preexisting soliton lattice corresponds to an almost
sinusoidal phase, and manifests itself as the pure shift of the
CDW peak without secondary satellites.
At higher applied currents however, a strong decrease
 of elastic modulus $C$ (corresponding to a decreasing $l$) leads to
non-harmonic modulation of the phase $\varphi$ and to higher-order satellites at $2\pi/L$
as in Fig.\ref{figure2}e.
 The strong variation of $C$ necessary to get a non-harmonic modulation 
 could be due to a screening effect by
 free carriers.
 In any case, the existence of the soliton lattice allows the electronic crystal to approach 
the commensurate modulation. In this work however, we did not observed any clear shift
of the 2$k_F$ reflection toward $0.75{\bf b^*}$ under current.
Note that this mechanism is based on the interaction between 
the CDW and the underlying lattice, without
taking explicitly into account the presence of extrinsic defects\cite{aubry}.

Another explanation of the secondary satellites could be the presence of an array of CDW edge-dislocations. A single CDW dislocation has already been observed in the bulk in\cite{lebolloch1}.
They can be formed
below the surface due to the charge transfer or other types of
stress near the surface. 
In the sliding state but for moderate electric field, the randomly distributed dislocations can be still pinned, leading to a mere change of the $2k_F$ profiles\cite{kirova}. At
high electric fields, the CDW dislocations can be depinned and form a periodic structure. Correlated dislocations of the host lattice are known in crystals stressed by the epitaxial layers (see for example \cite{epitaxy}).
The above model corresponds also to a non-harmonic modulation
and leads to higher-order satellites and to an asymmetric profile.

As already mentioned in the introduction, the secondary satellites
have been observed several times.
We never observed speckles but smooth profiles, 
despite the coherence properties of our x-ray beam.
The absence of any speckle is a strong indication that this phenomenon
could also have a dynamic origin. The secondary satellites could be
the consequence of the presence of a propagating mode, associated with the sliding of the condensate. 
An estimate of the frequency of the mode without current
can be obtained by extrapolating, at $\delta q_s$, the phason dispersion 
curve measured by inelastic neutron scattering\cite{hennion}.
For $\delta q_s=4.9\ 10^{-4}$ in ${\bf b^*}$ units, one obtains frequencies in the GigaHertz range (15 GHz at 175K and approximately 4 times larger at 75K),
i.e. a particulary low frequency mode. In the sliding state, a softening of this phason mode at $\delta q_s$
could explain the observed data since the diffracted intensity is inversely proportional to
the frequency squared of the vibration mode. The origin of the very soft mode remains to be clarified.

In conclusion, electronic correlations up to micrometer scale 
are observed, along the chain axis, in the sliding regime
of the blue bronze by using coherent x-ray diffraction. 
However, the appearance of super long-range correlations induced by CDW motion is still difficult to understand. Several scenarios are presented, based on the ordering of topological defects (solitons, edge-dislocations) or the softening of a phason mode. Additional experiments are necessary to clarify the origin of this phenomenon.

The authors would like to acknowledge the ID01 beamline staff at the ESRF,
especially C. Mocuta and T. Metzger, and P. Van den Linden  
for technical support; J.P. Pouget, M. Marsi and S. Brazovskii for fruitful discussions.
N. Kirova acknowledges the financial support of INTAS No 05-10000008-7972.
Part of this work was supported by ANR grant LoMaCoCup.


\begin{thebibliography}{1}
\bibitem{note7} The scheme of the reciprocal lattice can be found in our previous article\cite{lebolloch1}.
\bibitem{note6} Memory effects are well-known in blue bronze. 
A virgin state is obtained by increasing the temperature above $T_c$ and by cooling down without current.
\bibitem{note4} We have also observed that the whole 2$k_F$ profile was shifted 
 along ${\bf b^*}$ for some current values. 
We consider that this global shift is mainly due to the 
the finite reproducibility of the diffractometer since  
 the $(6,0,\overline{3})$ and the 2$k_F$ reflections have been measured 
successively after each current variation. We can estimate that the shifts of the 2$k_F$ profile with respect to
the current, if any, remain less than $\delta q=10^{-3}\AA^{-1}\sim 2.5\times\delta qs$.
\end{thebibliography}
\end{document}